# Stacking Order Driven Optical Properties and Carrier Dynamics in ReS$_2$

*Yongjian Zhou, Nikhilesh Maity, Amritesh Rai, Rinkle Juneja, Xianghai Meng, Anupam Roy, Xiaochuan Xu, Jung-Fu Lin, Sanjay Banerjee, Abhishek K. Singh\*, Yaguo Wang\**

Y. Zhou, X. Meng, Prof. Y. Wang
Department of Mechanical Engineering
The University of Texas at Austin
Austin, TX 78712, USA
E-mail: yaguo.wang@austin.utexas.edu

N. Maity, R. Juneja, Prof. A. K. Singh
Materials Research Centre
Indian Institute of Science, Bangalore-560012
E-mail: abhishek@iisc.ac.in

A. Rai, A. Roy, Prof. S. K. Banerjee
Microelectronics Research Center
The University of Texas at Austin
Austin, Texas 78758 USA

Prof. X. Xu,
State Key Laboratory on Tunable Laser Technology,
Harbin Institute of Technology, Shenzhen,
Shenzhen, Guangdong 518055, China;

Prof. J.-F. Lin, Prof. Y. Wang
Texas Materials Institute
The University of Texas at Austin
Austin, TX 78712, USA

Prof. J.-F. Lin
Department of Geological Sciences
Jackson School of Geosciences
The University of Texas at Austin
Austin, TX 78712, USA

**Abstract text**

Two distinct stacking orders in ReS$_2$ are identified without ambiguity and their influence on vibrational, optical properties and carrier dynamics are investigated. With atomic resolution scanning transmission electron microscopy (STEM), two stacking orders are determined as AA stacking with negligible displacement across layers, and AB stacking with about a one-unit cell displacement along the *a* axis. First-principle calculations confirm that these two stacking orders correspond to two local energy minima. Raman spectra inform a consistent difference of modes I & III, about 13 cm$^{-1}$ for AA stacking, and 20 cm$^{-1}$ for AB stacking, making a simple tool for determining the stacking orders in ReS$_2$. Polarized photoluminescence (PL) reveals that AB stacking possesses blue-shifted PL peak positions, and broader peak widths, compared with AA stacking, indicating stronger interlayer interaction. Transient transmission measured with femtosecond pump probe spectroscopy suggests exciton dynamics being more anisotropic in AB stacking, where excited state absorption related to Exc. III mode disappears when probe polarization aligns perpendicular to *b* axis. Our findings underscore the stacking-order driven optical properties and carrier dynamics of ReS$_2$, mediate many seemingly contradictory results in literature, and open up an opportunity to engineer electronic devices with new functionalities by manipulating the stacking order.

**Main Text**

Transition metal dichalcogenides (TMDs) represent a rich family of 2D semiconductors with many intriguing quantum phenomena for novel electronic devices.[1-3] ReS$_2$, a rising star among TMDs, has drawn much attention in recent years. ReS$_2$ possesses a distorted 1T triclinic crystal structure where the additional d valence electrons of Re atoms form zigzag Re chains parallel to the *b* axis, drastically reducing its symmetry. Even though properties of bulk ReS$_2$ have been studied since 1997,[4-15] research on the 2D form of ReS$_2$ only began to surge around 2014.[16] Comparing with other TMDs, interlayer coupling of ReS$_2$ is much weaker.[16] The uniqueness of ReS$_2$ lies in its in-plane anisotropic properties, which have been demonstrated as early as 2001 in bulk.[9] In 2D ReS$_2$, properties observed are polarization-dependent excitons,[17, 18] non-linear absorption,[19] electron transport and SHG emission,[20, 21] *etc*. Compared with black phosphorous (BP), which also shows in-plane anisotropic properties, ReS$_2$ is more stable in ambient environments, which makes it more suitable for optoelectronic devices. Among the studies of ReS$_2$, many contradictory findings have been reported.[18, 22-27] For example, Tongay *et al.* suggested that the monolayer behavior of ReS$_2$ still persists in bulk,[16] which, however, was challenged by several later studies.[22, 24, 25] Also, scanning transmission electron microscope (STEM) images of multilayer ReS$_2$ show drastically different features reported by different groups.[25, 26] Similar inconsistencies also exist in the determination of indirect-direct bandgap transition,[18] Raman vibrational modes and second harmonic generation (SHG) spectra.[20-24, 28] These results suggest that some other intrinsic parameters governing the electronic and optical properties of multilayer ReS$_2$ are not well understood. In trilayer graphene, researchers found that the two most stable stacking orders of ABC and ABA display dramatically different behavior in electronic screening of external field,[29] Raman vibrations,[30] and even electron transport.[31] In TMDs, however, the effect of stacking order was mainly studied with first principles calculations, and suggest stacking-order driven valence band splitting and exciton binding energy change.[32] Only a couple of experimental studies investigated the stacking order in TMDs, mainly using TEM, Raman or PL.[33, 34] Possible stacking order in ReS$_2$ has not been identified, even though some Raman studies indicated the existence of different stacking orders.[22-24]

In this work, first we identified two distinct stacking orders (AA & AB) of multilayer ReS$_2$ with atomic resolution STEM without any ambiguity. Similar to the graphene,[29] individual layers of AA stacking sit on top of each other, while for AB stacking a shift along *a* axis occurs. First principles calculations with density functional theory (DFT) confirmed that these two stacking orders correspond to two local energy minima. Second, we studied the stacking-order driven vibrational, optical properties and exciton dynamics with Raman, PL and femtosecond pump-probe spectroscopy. Raman spectra show larger differences between mode III and I for AB stacking. PL spectra revealed higher exciton peak positions and smaller exciton lifetimes for AB stacking. Both Raman and PL results suggested stronger interlayer interaction in AB stacking. Pump-probe results revealed excited excitons are weakened in AB stacking, especially when probe polarization is perpendicular to *b* axis, which causes the dominance of phase-filling effects. Our experimental results are not only consistent with DFT calculations, but also explain well the inconsistency of previously reported by considering stacking order. This study represents a significant step forward in understanding and manipulating the properties of ReS$_2$, which could facilitate its application in next-generation electronic and nonlinear optical devices.

Samples are prepared by mechanical exfoliation, with single crystals purchased from HQ Graphene. Naturally, ReS$_2$ has cleaved edges along b axis, which are confirmed by measuring mode V intensity with polarized Raman, with parallel incident and collection light (532 nm) (Figure S7, Supporting Information). All samples under study show a well-defined *b* axis. To determine the stacking order unambiguously, we conducted STEM measurements on two samples. Based on Raman and PL, sample AA has about 4 layers, and about 2 layers for sample AB.[22] The STEM images of samples AA and AB (**Figure 1**a and 1b) demonstrate completely different features. However, both of them have well-defined Re-Re chain, suggesting no rotational displacement among layers. Figure 1a shows a simple arrangement of Re atoms, which is very similar to the published STEM images on monolayer ReS$_2$.[26] Sample AB shows a rather complicated pattern (Figure 1b), with periodic bright spots and bi-atom pairs. On top of the STEM images, the proposed displacement between two layers is sketched, with the almost overlapping atoms marked in red dashed circles. For sample AA, it is reasonable to perceive that there is almost no displacement among layers. For sample AB, the bright spots are possibly the positions where atoms from different layers almost overlap. What appear to be bi-atom pairs are possibly those atoms from different layers that sit close to each other. This feature can be reproduced if the top layer moves relative to the bottom layer *a* axis about one unit cell along.

To confirm our proposed structure of stackings AA and AB, we also performed ab initio calculations for bilayer ReS$_2$. As shown in Figure 1c, after scanning through the displacement space along *a* axis, two local energy minima occur. Figure 1d & 1e depict the relaxed lattice structures at these two minima. At first energy minima, the two layers have minimal displacement. At the second energy minima, there is about a one-unit cell (~2.5 Å) displacement along *a* axis. This is consistent with the STEM images and confirms our proposed structures for both stacking AA and AB. Qiao *et al.* proposed two possible stacking orders in $ReS_2$ based on the vibrational states measured with Raman and DFT calculations, where the second layer has a rotational displacement of 60° and 120° with respect to the first layer.[22] However, the problem with this proposed structure is that the universal *b* axis (Re-Re chain) should no longer exist, which contradicts the previously published findings of a clearly defined *b* axis for $ReS_2$.[26] With similar approaches, He *et al.* proposed three possible stacking orders with translational displacements along the direction perpendicular to *b* axis.[24] Compared with these obscure results, our STEM images and ab initio calculations show without ambiguity that there are two possible stacking orders existing in $ReS_2$, AA stacking (with negligible displacement among layers), and AB stacking (with translational displacement along *a* axis). In order to reveal the impact of stacking order on vibrational properties, Raman measurements were conducted on multiple exfoliated samples, with thicknesses ranging from 55 nm to 580 nm (Figure S6). **Figures 2**a & b show mode I and mode III of all samples, which correspond to the $A_g$- (cross-plane) and $E_g$-like (in-plane) modes. Among the 9 samples studied, the peak of mode III lies consistently slightly above 150 $cm^{-1}$, while that of mode I shows up either close to 130 $cm^{-1}$ or 140 $cm^{-1}$. When plotting the difference between these two modes, Δ = mode III – mode I, the samples may be clearly categorized into two groups (A & B) (Figure 2c). Group B has a Δ of about 20 $cm^{-1}$, and group A has a Δ of about 13 $cm^{-1}$. We also calculated the Raman spectra of stacking AA and AB, as plotted in Figure 2d. What we observed is that mode III peak positions are the same in both stackings, but the mode I peak in stacking AA is about 4.7 $cm^{-1}$ higher than that in stacking AB. These calculations confirm that group A samples in Figure 2c possess stacking

WILEY-VCHWILEY-VCHorder AA and group B samples possess stacking order AB. Even more astonishing, the stacking order of ReS$_2$ is robust enough to persist even in bulk samples. The difference between Raman peaks of mode I and mode III can be used as an indicator of stacking order in ReS$_2$, similar to what was proposed by Qiao *et al.* in multilayer ReS$_2$[22]. It is not surprising that mode III does not vary much among different samples since it corresponds to in-plane lattice vibrations. Mode I originates from out-of-plane vibrations within a single layer, and is more prone to the stacking order. Higher mode I frequency in stacking AA samples indicates stronger interatomic bonding within a single layer along cross-plane direction, which may be further understood as showing weaker interlayer interaction. We also calculated the Raman spectra of interlayer breathing (*B*) and shear modes (S$_{\parallel}$, S$_{\perp}$), as shown in Figure 2e. S$_{\parallel}$ represents the shear mode along Re chain (***b*** axis) and S$_{\perp}$ represents the shear mode perpendicular to the Re chain. For all three low-frequency modes, the values in stacking AB are higher than those in stacking AA, with $\Delta S_{\parallel}$ = 3.1 cm$^{-1}$, $\Delta S_{\perp}$ = 0.61 cm$^{-1}$, $\Delta B$ = 0.66 cm$^{-1}$. Higher-frequency breathing and shear modes in stacking AB indicate stronger interlayer interaction.[35] In the relaxed crystal structure, the interlayer distance of stacking AB is 2.59 Å, smaller than that of stacking AA, 2.71Å, which also supports the stronger interlayer interaction in stacking AB.

To investigate how the stacking order of ReS$_2$ affects its optical properties, we conducted polarized PL measurements. The results of two representative samples, AA* (sample 2 in Figure 2a) and AB* (sample 7 in Figure 2b), are plotted in **Figure 3**. We found that while the PL spectra do not show obvious affect from the polarization of incident beam, they are very sensitive to that of collection beam. As a result, we fixed the polarization of incident beam along ***b*** axis (0°), but, aligned the polarization of collection beam both along (0°) and perpendicular (90°) to ***b*** axis. Voigt function was used to fit the PL spectra to determine the peak positions and widths (full width at half maximum, FWHM), also plotted in Figure 3.

666

Four peaks are identified, with the peak around 1.4 eV being the optical transition at the indirect band gap, and the other three peaks assigned to exciton I, II and III, respectively.[18] For both samples, Exc. I is more prominent when collection polarization is along 0°, and Exc. II is more prominent at 90°. This is consistent with previous studies suggesting that Exc. I and Exc. II are polarized along different directions.[18] For Exc. III, in both stackings it is more prominent along 0° than 90°. However, the PL intensity of Exc. III in sample AB* is much weaker than that in AA*. For AB* Exc. III almost vanishes at 90°. In ReS$_2$, the Rydberg series was observed and the behavior can be well-explained by Wannier excitons.[17] Exc. III was assigned as an excited Rydberg exciton state of lower-lying excitons.[17, 18] Since the excited states of Exc. I & II lie very close to each other and both close to the Exc. III energy, even though they have different preferable polarization, we can view Exc. III as a superposition of two excited states from Exc. I & II. The higher PL intensity of Exc. III along 0° collection polarization suggests that the excited state of Exc. I contributes a larger portion to Exc. III. High lying excited states such as Exc. III are very sensitive to the environment.[36] The greatly reduced PL intensity of Exc. III in AB* indicates that the excited excitons are weakened substantially, especially along 90° collection polarization.

For all three excitons, AB* has blue-shifted PL peak positions (20~25meV) and broader widths than stacking AA*. Even though only two samples are presented in Figure 3, this general trend is seen with other samples sharing the same stacking order (Figure S8-S16, Supporting Information). The position of the PL exciton peak, $E_{exc}$, is determined by both the optical transition of electronic band gap ($E_{BG}$) and exciton binding energy ($\Delta_{binding}$): $E_{exc} = E_{BG} - \Delta_{binding}$. Broader exciton peak width in AB* suggests a shorter radiative lifetime, which further implies stronger exciton oscillation strength.[37] Our ab initio calculations predicted interlayer distances of 2.71 Å for AA stacking and 2.59 Å for AB stacking. A shorter interlayer distance allows the wave functions of carriers in different layers to have a higher probability to interact, which results in a larger dielectric constant and stronger interlayer interaction. It is known that a larger dielectric constant environment screens the Coulomb force of intralayer excitons, which usually results in blue shift of exciton.[38] Stronger interlayer interaction means stronger carrier scattering, and a shorter radiative lifetime.

To study how the stacking order affects the exciton dynamics, we performed ultrafast pump-probe spectroscopy with degenerate pump/probe pulses at 790 nm. Since the photon energy of 1.57 eV resonates with Exc. III, the transient transmission signals mainly relate to the dynamics of Exc. III. **Figure 4** shows the transient transmission signals for AA* and AB*. Because the signals do not have obvious dependence on pump polarization, but are very sensitive to probe polarization (Figure S18, Supporting Information), only four polarization combinations are presented. We chose cross-polarized pump and probe beams for ease of data acquisition. For AA*, the transmission signals show negative peaks for both 0° and 90° probe polarizations. For AB*, the signal displays a negative peak when the probe is along 0°, while a very sharp positive peak when the probe is along 90°.

Under resonant probe conditions, the transmission signal is usually related to the imaginary part of the refractive index,[39-42] and hence reflects the absorption change in the material. A decrease of transmission (negative peaks) at time zero indicates increasing absorption of probe photons, while positive peaks indicate decreasing absorption. Since pump pulse excites carriers to the Exc. III level, the negative peak means these excited carriers can further absorb probe photons and experience intraband transition to even higher energy levels (Figure 4a~c). This phenomenon is called excited state absorption (ESA). To observe ESA experimentally, the laser pulse width has to be shorter than the relaxation time of Exc. III. The pulse width of our femtosecond laser is about 200 fs (FWHM) at the sample position, much shorter than the reported exciton recombination time in $ReS_2$.[43] The positive $\Delta T/T$ peak (Figure 4d) is usually explained as a result of phase-filling. Due to fermion nature of electrons (or holes), each

8probability to interact, which results in a larger dielectric constant and stronger interlayer interaction. It is known that a larger dielectric constant environment screens the Coulomb force of intralayer excitons, which usually results in blue shift of exciton.[38] Stronger interlayer interaction means stronger carrier scattering, and a shorter radiative lifetime.

To study how the stacking order affects the exciton dynamics, we performed ultrafast pump-probe spectroscopy with degenerate pump/probe pulses at 790 nm. Since the photon energy of 1.57 eV resonates with Exc. III, the transient transmission signals mainly relate to the dynamics of Exc. III. **Figure 4** shows the transient transmission signals for AA* and AB*. Because the signals do not have obvious dependence on pump polarization, but are very sensitive to probe polarization (Figure S18, Supporting Information), only four polarization combinations are presented. We chose cross-polarized pump and probe beams for ease of data acquisition. For AA*, the transmission signals show negative peaks for both 0° and 90° probe polarizations. For AB*, the signal displays a negative peak when the probe is along 0°, while a very sharp positive peak when the probe is along 90°.

Under resonant probe conditions, the transmission signal is usually related to the imaginary part of the refractive index,[39-42] and hence reflects the absorption change in the material. A decrease of transmission (negative peaks) at time zero indicates increasing absorption of probe photons, while positive peaks indicate decreasing absorption. Since pump pulse excites carriers to the Exc. III level, the negative peak means these excited carriers can further absorb probe photons and experience intraband transition to even higher energy levels (Figure 4a~c). This phenomenon is called excited state absorption (ESA). To observe ESA experimentally, the laser pulse width has to be shorter than the relaxation time of Exc. III. The pulse width of our femtosecond laser is about 200 fs (FWHM) at the sample position, much shorter than the reported exciton recombination time in $ReS_2$.[43] The positive $\Delta T/T$ peak (Figure 4d) is usually explained as a result of phase-filling. Due to fermion nature of electrons (or holes), each quantum state on the same energy level only allows for two electrons with opposite spins. When all the states are occupied, the Pauli-blocking effect prevents further excitation of carriers and hence absorption decreases. In light of this understanding, we propose that for AA*, ESA dominates near time zero for both probe polarizations: for AB*, ESA only dominates when the probe is along 0°, while the phase-filling effect dominates for 90° probe. The phenomena observed in transient transmission experiments are consistent with the non-linear absorption results measured in $ReS_2$ (Figure S19, Supporting Information). In stacking AA, strong ESA exists with light polarization along both 0° and 90°. In stacking AB, ESA dominates when light polarization is 0°, and saturable absorption (SA, where absorption decreases with light intensity by the phase filling effect) dominates when polarization is near 100°.[19] Our nonlinear absorption data of AB stacking was published in 2018, before which time the existence of stacking order in $ReS_2$ had not yet been recognized in the scientific community.

ESA requires a large number of excitons to exist at level Exc. III, which in turn requires adequate available states at Exc. III. For stacking AA, since the relative shift between layers is negligible, it is reasonable to expect that these higher exciton levels are preserved along all directions, similar to those of monolayer. In stacking AB however, the one unit-cell relative shift along *a* axis causes disorder to be introduced along the perpendicular direction, as a result the high lying exciton levels may be disrupted and fewer states available at Exc. III. This explanation is also consistent with the PL spectra presented in Figure 3 for both collection polarizations of AA*, and for the 0º collection in AB*, where the PL peaks of Exc. III are quite prominent. At the 90º collection in AB*, the PL peak of Exc. III almost disappears. This weak PL peak indicates less available states, and therefore we expect to observe stronger phase-filling effects in transient absorption and SA in nonlinear absorption.

The decaying processes following the peaks correspond to several factors including the relaxation from high-lying to lower exciton levels, indirect band edge, and exciton recombination. For the 90° probe polarization in stacking AB* a sign change appears. In CVD grown TMDs, such as $MoSe_2$,[44] the sign change is usually explained as carrier trapping/releasing by the mid-gap defect states, followed by carrier recombination. Since all of our $ReS_2$ samples are mechanically exfoliated and the sign change only appears in the 90º probe polarization case, such sign change should not be associated with mid-gap defect states. One possible explanation rather is a rapid scattering of carriers to other polarizations, most probably 0°. As mentioned earlier, Exc. III can be viewed as a superposition of Exc. I and Exc. II. For both AA* and stacking AB*, the PL spectra show stronger peaks of Exc. I when the collection polarization aligns along *b* axis, and stronger peaks of Exc. II when collection polarization aligns along 90° (perpendicular to *b* axis). The excitons at Exc. III level along 90° could quickly scatter to those along 0°, emptying some states for further absorption of 90º probe photons. This process shows some similarity with momentum randomization in excited carriers, with comparable scattering times (~several hundred fs).[45]

Identifying the stacking order of $ReS_2$ mediates many seemingly "controversial" results published in literature. For example, the STEM images of multilayer $ReS_2$ reported by different groups are inconsistent, some having very clean and well-arranged Re atoms,[18] and others not.[25] These different images must come from samples with different stacking orders. Second, Aslan *et al.* reported that $ReS_2$ maintains indirect band semiconductors even in monolayer and without signature transition,[18] while Mathias et al. observed that monolayer and bilayer $ReS_2$ have direct band gaps.[46] A third example is the SHG spectra, as reported by Song *et al.* where SHG signals of $ReS_2$ can only be observed in even number layers,[20] while Dhakal *et al.* showed that the SHG signal increases with thickness of $ReS_2$ starting from monolayer, regardless of the number of layers.[21] All of these discrepancies are explained with ease if the factor of stacking order is considered. SHG is determined by the symmetry of crystal. With AA stacking, where there is negligible displacement among layers, the symmetry is expected to be the same as the monolayer, which explains the observation by Song *et al.* With AB stacking, symmetry is defined by two-layers, rather than one, hence the SHG signal only appears in even numbers of layers. Furthermore, our results underscore the stacking-order driven optical properties and carrier dynamics in $ReS_2$, which opens opportunities to pursue new functionalities and engineer new electronic devices by manipulating stacking order. While surveying 11 exfoliated $ReS_2$ samples, we did observe 2 samples showing mixed stacking orders, as shown in Figure S20, which might warrant further study.

**Experimental Section**

*Sample Preparation:* Samples were prepared by mechanical exfoliation, with single crystals purchased from HQ Graphene. For STEM measurements, the samples were firstly exfoliated onto $SiO_2$/Si (Addison Engineering). Then, after being characterized by Raman and PL, the few-layer samples were transferred onto the Quantifoil TEM grid (TED PELLA, INC). For Ultrafast carrier dynamics measurements, the samples were exfoliated onto glass substrate (TED PELLA, INC).

*Raman and PL Measurements:* Both Polarized Raman and PL measurements were conducted by inVia™ confocal Raman microscope (Renishaw) with 532 nm excitation. For Raman, 2400 lines $mm^{-1}$ grating were used. In order to control the polarization, the samples were mounted on a rotational stage. For PL, 1200 lines $mm^{-1}$ grating were used.

*STEM measurements:* Experiments were conducted by JEOL JEM-ARM 200F at 80kV.

*Ultrafast carrier dynamics measurements:* The ultrafast carrier dynamics measurements were made by degenerate pump probe method at room temperature. Both pulses (pump and probe) were generated from a femtosecond Ti: Sapphire Oscillator (spectra physics, Tsunami), with 20 nm spectral linewidth (FWHM) and 200fs pulse width at the sample position. The spot size ($1/e^2$ diameter of Gaussian beam) for pump and probe were about 13.3 μm and 6.7 μm, respectively.

**Supporting Information**
Supporting Information is available from the Wiley Online Library or from the author.

**Acknowledgements**
The authors are grateful for the supports from National Science Foundation (NASCENT, Grant No. EEC-1160494; CAREER, Grant No. CBET-1351881, CBET-1707080, Center for Dynamics and Control of Materials DMR-1720595);  AKS, NM and RJ thank the Materials Research Centre and Supercomputer Education and Research Centre of Indian Institute of Science for providing computing facilities. ARai, ARoy and SKB thank the support by the National Science Foundation National Nanotechnology Coordinated Infrastructure grant (NNCI-1542159).

Received: ((will be filled in by the editorial staff))
Revised: ((will be filled in by the editorial staff))
Published online: ((will be filled in by the editorial staff))

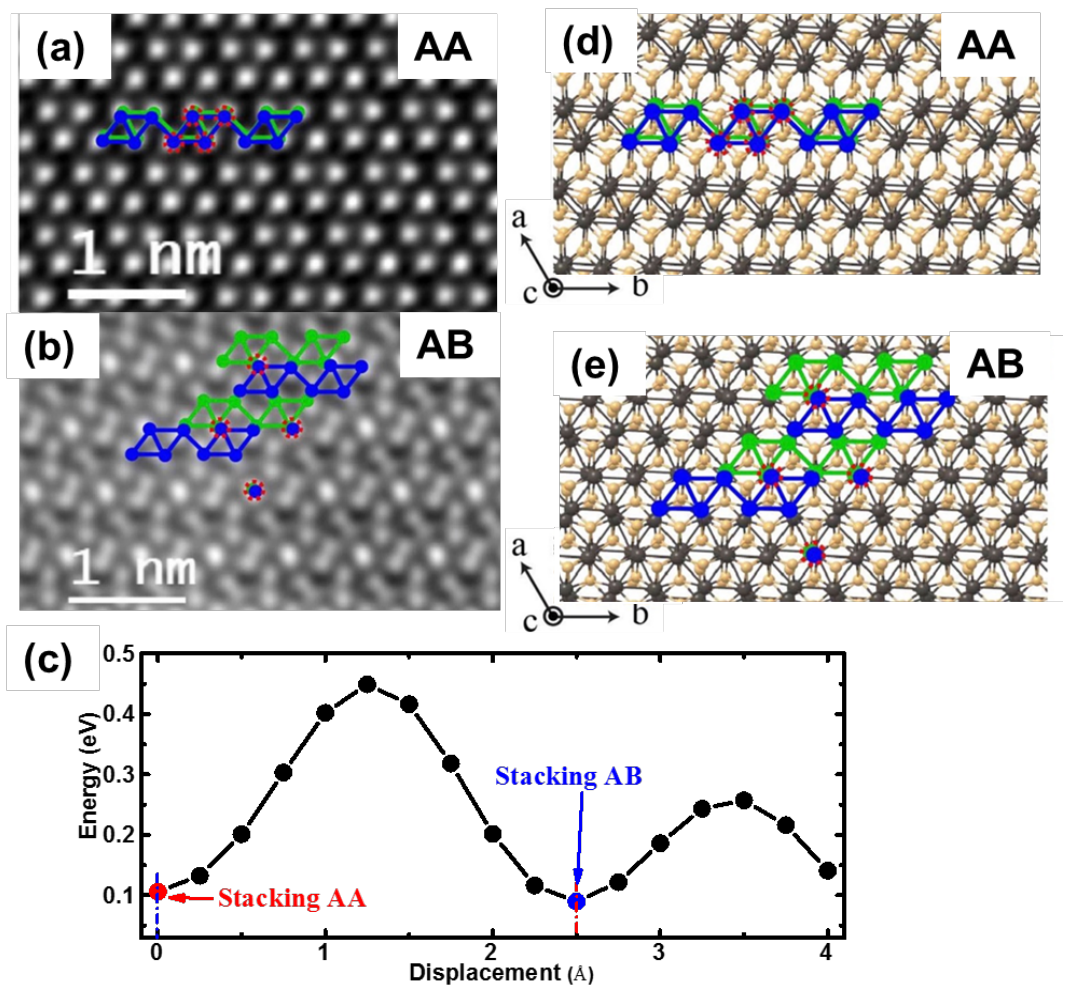

**Figure 1.** (a) & (b) STEM images of AA and AB stacking orders, with proposed relative replacement between two layers. Green color represents the bottom layer, blue the top layer. The red dashed circles mark the bright dots in STEM images, which come from the almost overlapping atoms of two layers. (c) First-principles calculations of total energy of a bilayer ReS$_2$ structure against displacement along *a* axis. (d) & (e) Snapshots of crystal structures predicted with ab-initio calculations at two local energy minima, as they correspond to AA and AB stacking orders. Note the relative shift in AB stacking along *a* axis, which is negligible in AA stacking.

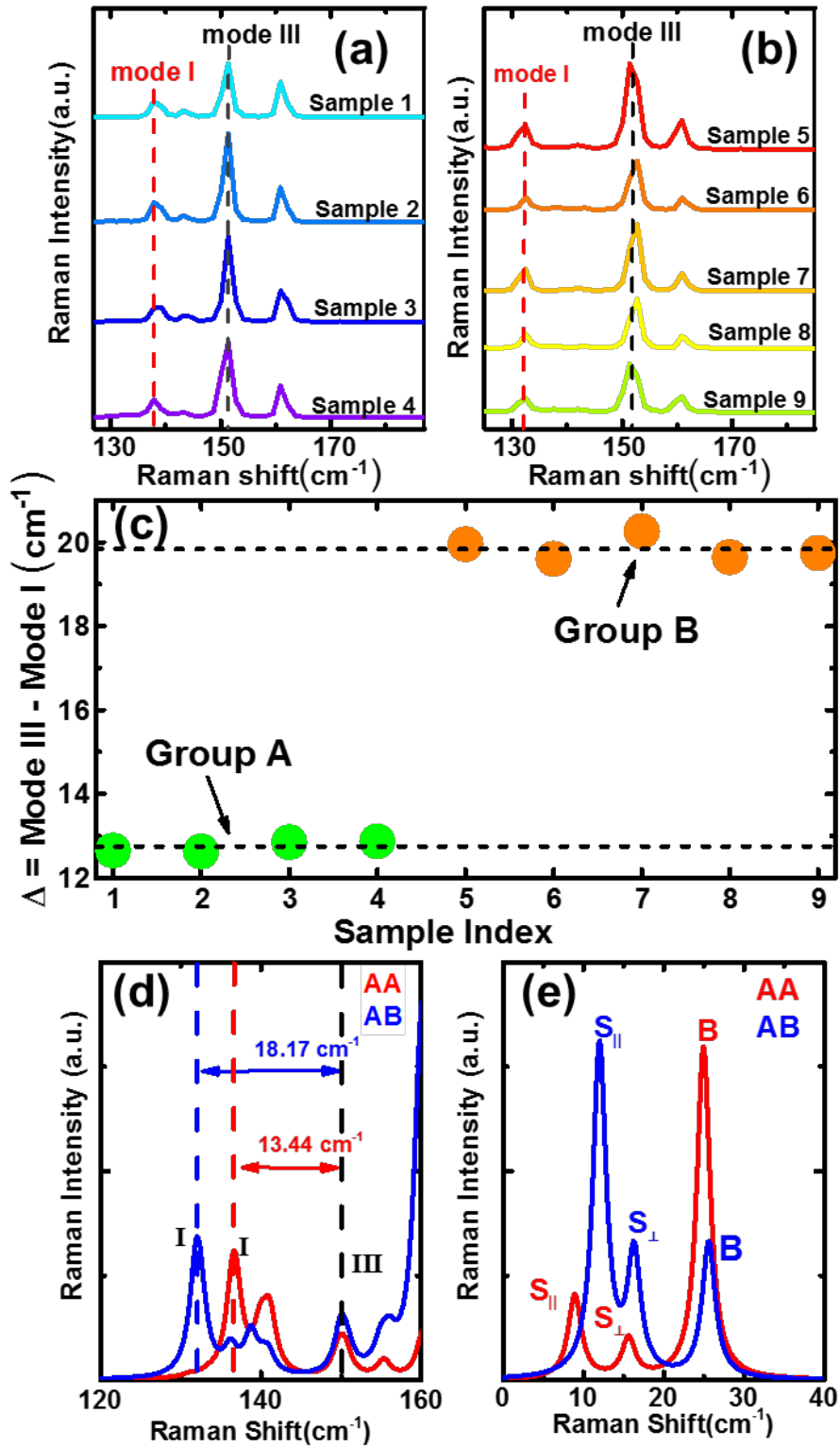

**Figure 2.** Raman spectra of mode I and mode III in 9 bulk ReS$_2$ samples: (a) group A and (b) group B. (c) The Δ of mode III and mode I, which clearly shows two different groups corresponding to AA (group A) and AB (group B) stacking. (d) Calculated Raman spectra of mode I and mode III in both stackings. (e) Calculated Raman spectra of low frequency breathing (B) and shear modes (S$_∥$, S$_⊥$) in both stackings.

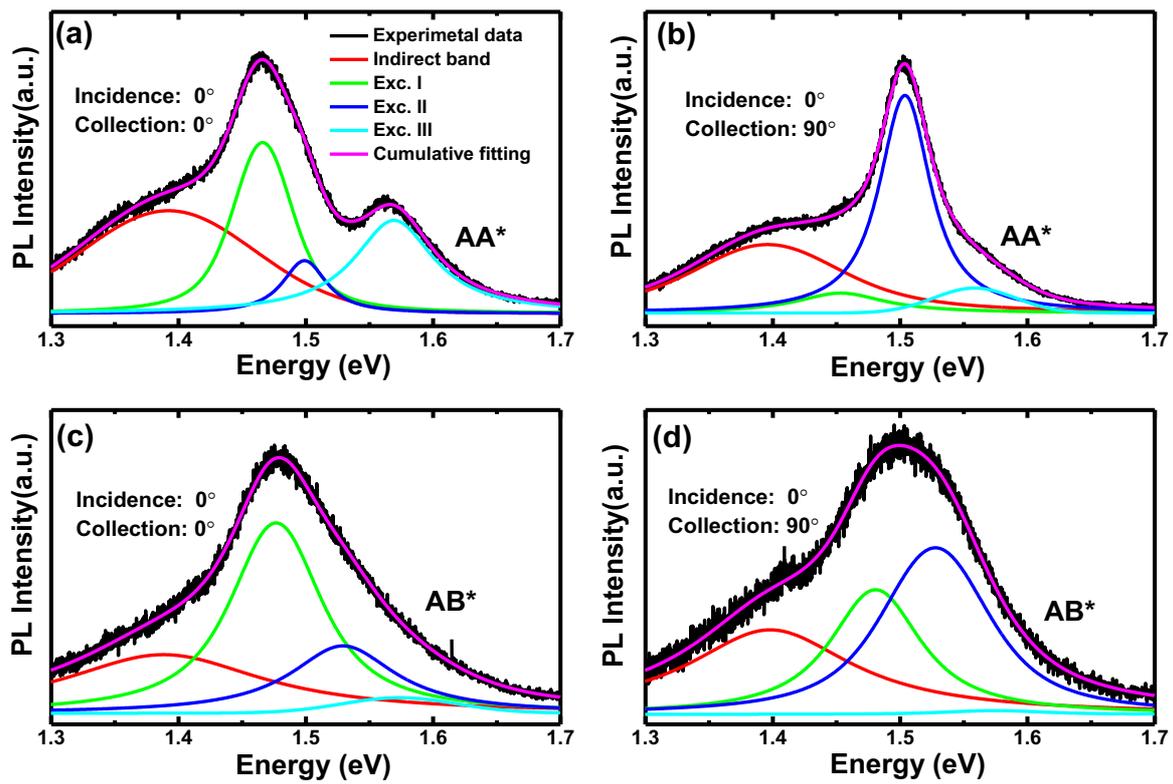

**Figure 3.** (a) PL spectra of sample AA* with 0 ° polarization for both incident and collection beams; (b) PL spectra of sample AA* with 0 ° polarization for incident beam and 90 ° for collection beam; (c) PL spectra of sample AB* with 0 ° polarization for both incident and collection beams; (d) PL spectra of sample AB* with 0 ° polarization for incident beam and 90 ° for collection beam.

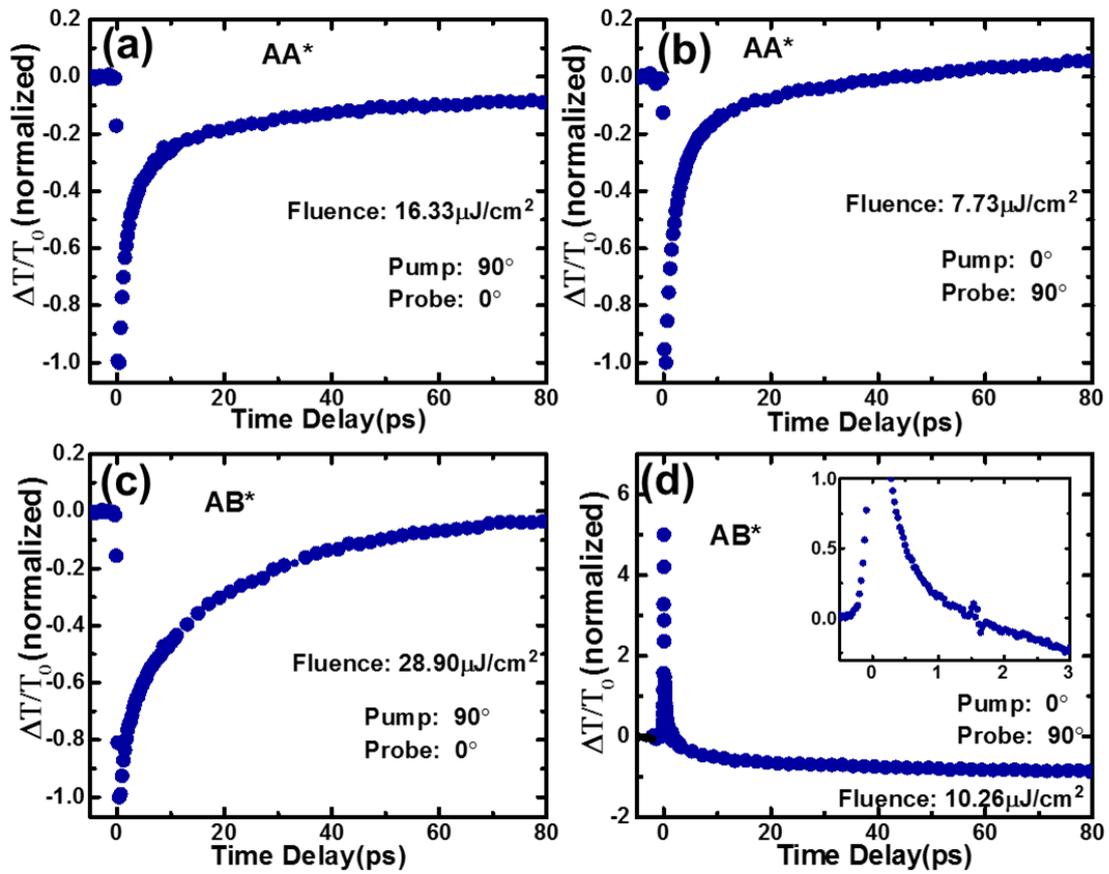

**Figure 4.** Transient transmission signal measured with femtosecond pump probe spectroscopy: in stacking AA* (a) & (b), with pump polarized at 90º and probe at 0º (a); and with pump polarized at 0º and probe at 90º (b); and in stacking AB* (c) & (d), with pump polarized at 90º and probe at 0º (c); and with pump polarized at 0º and probe at 90º (d).

**Table of Contents**
Two stacking orders of $ReS_2$ are identified. Stacking AA has negligible displacement across layers and stacking AB has a one-unit cell displacement along a-axis. AB stacking has stronger interlayer coupling than AA. The cross-layer displacement in AB stacking disrupts excited state excitons. Vibrational, optical properties and carrier dynamics in two stacking orders are drastically different.

**Keyword** 2D materials, $ReS_2$, stacking order, Scanning transmission electron microscopy, First-principles Calculations

*Yongjian Zhou, Nikhilesh Maity, Amritesh Rai, Rinkle Juneja, Xianghai Meng, Anupam Roy, Xiaochuan Xu, Jung-Fu Lin, Sanjay Banerjee, Abhishek K. Singh\*, Yaguo Wang\**

**Stacking Order Driven Optical Properties and Carrier Dynamics in $ReS_2$**

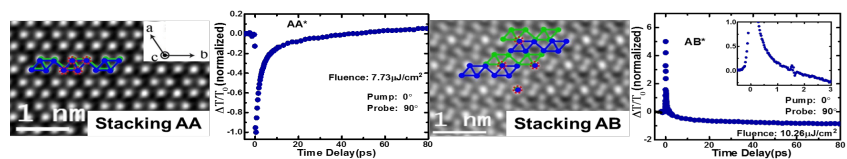